\documentclass[Journal,NoLists,NoLineNumbers]{ascelike-new}

\usepackage[utf8]{inputenc}
\usepackage[T1]{fontenc}
\usepackage{lmodern}
\usepackage{graphicx}
\usepackage[figurename=Fig.,labelfont=bf,labelsep=period]{caption}
\usepackage{color,soul}
\usepackage[colorlinks=true,citecolor=blue,linkcolor=black,urlcolor=blue]{hyperref}
\usepackage[open,openlevel=1]{bookmark}

\NameTag{}

\begin{document}

\title{Design principles of multi-map variation in biological systems}

\author[1]{Juan F. Poyatos}

\affil[1]{Logic of Genomic Systems Lab (CNB-CSIC) Madrid. Email: jpoyatos@cnb.csic.es}

\maketitle

% Please include an abstract:
\begin{abstract}

Complexity in biology is often described using a multi-map hierarchical architecture, where the genotype, representing the encoded information, is mapped to the functional level, known as the phenotype, which is then connected to a latent phenotype we refer to as fitness. This underlying architecture governs the processes driving evolution. Furthermore, natural selection, along with other neutral forces, can, in turn, modify these maps. At each level, variation is observed. Here, I propose the need to establish principles that can aid in understanding the transformation of variation within this multi-map architecture. Specifically, I will introduce three, related to the presence of modulators, constraints, and the modular channeling of variation. By comprehending these design principles in various biological systems, we can gain better insights into the mechanisms underlying these maps and how they ultimately contribute to evolutionary dynamics.

%putatitve other phenotypic layers.

\end{abstract}

\newpage
%%%%%%%%%%%%%%%%%%%%%%%
\section{Introduction}
\begin{figure}[bt]
\centering
\includegraphics[width=12cm]{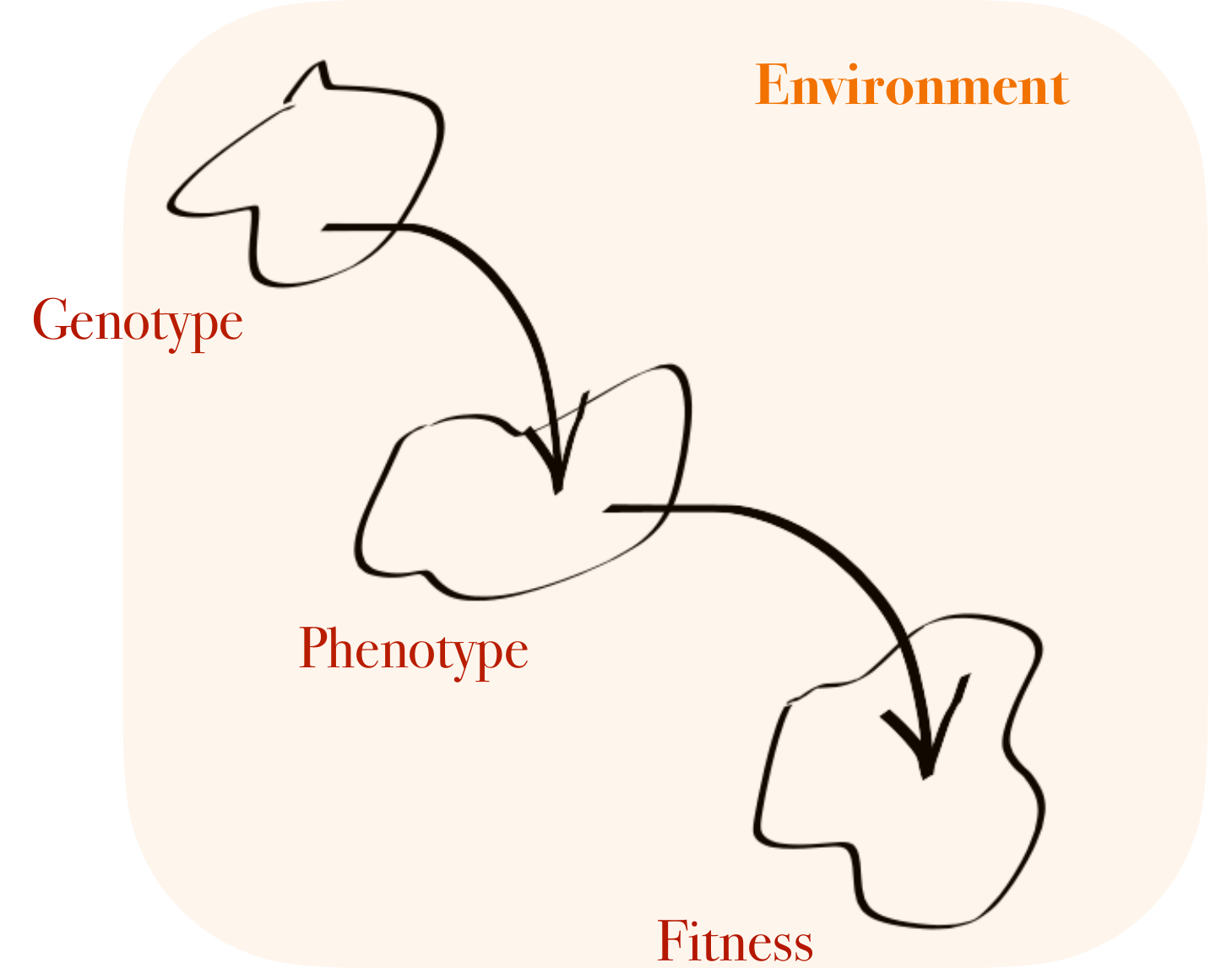}
\caption{Multi-map architecture of biological complexity. The multi-map architecture is a fundamental framework for understanding biological complexity. It involves the mapping of genotype to phenotype and that of phenotype to fitness. The action of these mappings can be shaped by environmental factors, altered by selection processes, and, in specific instances, the direction of the arrows can be reversed.}
\label{fig1}
\end{figure}

%%%%%%%%%%%%%%%%%%%%%%%
It is well recognized that biology encompasses numerous mappings, i.e., laws of transformation that are both dynamically and empirically sufficient~\cite{lewontin_genetic_1974}. In a given environment, we observe a mapping from genotype to phenotype, representing function, and from phenotype to fitness, which can be perceived as a latent phenotype that is not readily observable but connected to selection (Figure~\ref{fig1}). While it is acknowledged that some of these mappings may not be unidirectional, for the sake of simplicity, let's maintain that assumption.

Within this framework, we encounter variation at the genotypic, phenotypic, and fitness levels~\cite{hallgrimsson_variation_2005}. This variation associates with three crucial characteristics frequently discussed in biology: robustness, plasticity, and evolvability~\cite{waddington_strategy_1957}. These properties elucidate how variation at one level can either be reduced at a different level (robustness) or manifest as a response to perturbations (plasticity) at various levels. They also highlight how variation drives evolution (evolvability) and how environmental factors modify all these aspects.

Below, I present an argument supporting the existence of design principles governing the impact of variation in this multi-mapped architecture, which I will refer to as "multi-map variation" throughout this discussion. While the absolute generality of these principles may be debatable, I advance that they can be observed in various implementations in a wide range of biological contexts. Therefore, this work is part of the broader search for generalizable design principles in various areas of biology. The identification of specific principles and their biological implementations has played a crucial role, for example, in the development of a theory of biological networks~\cite{soyer_search_2012,alon_introduction_2019}. 

With this in mind, I propose three integral elements that contribute to what could eventually be understood as a general theory of multi-map variation. First, we observe the existence of "malleable" agents that control variation. Second, we encounter "hard" restrictions on variation, acting as limiting factors or constraints that prevent certain deviations. Third, we notice a clear "modular" channeling of variation.
Characterizing these principles in different biological systems will allow us to unravel the mechanisms underlying the mapping process, the forces that shape evolutionary trajectories, and the factors that influence the emergence of new traits. Ultimately, this knowledge can enhance our ability to interpret and predict evolutionary dynamics in diverse biological contexts. Let me explain these three elements further below.

%%%%%%%%%%%%%%%%%%%%%%%%%%%%%%%%%%%%%%%%%%%%%%%%%%%%%%%%%%%%%%%%%%%%%
%%%%%%%%%%%%%%%%%%%%%%%%%%%%%%%%%%%%%%%%%%%%%%%%%%%%%%%%%%%%%%%%%%%%%

\section{Principles}

%%%%%%%%%%%%%%%%%%%%%%%%%%%%%%%%%%%%%%%%%%%%%%%%%%%%
\subsection{“Malleable” agents of variation control}
%%%%%%%%%%%%%%%%%%%%%%%%%%%%%%%%%%%%%%%%%%%%%%%%%%%%

Imagine a scenario where a given biological system undergoes multiple mutations. In more technical terms, we can examine a trajectory of mutation accumulation (MA), where most mutations are allowed to accumulate at random over a given time period~\cite{halligan_spontaneous_2009}. This represents the fundamental variation at the genotype level. My argument is that there are elements within the system that modulate the phenotypic variation generated as a consequence of these mutations by exhibiting a consistent pattern of epistasis (interactions between these elements and the  mutations)~\cite{rutherford_hsp90_1998,bergman_evolutionary_2003,schell_modifiers_2016,poyatos_genetic_2020}. I refer to these units as "agents of variation control".

\begin{figure}[bt]
\centering
\includegraphics[width=16cm]{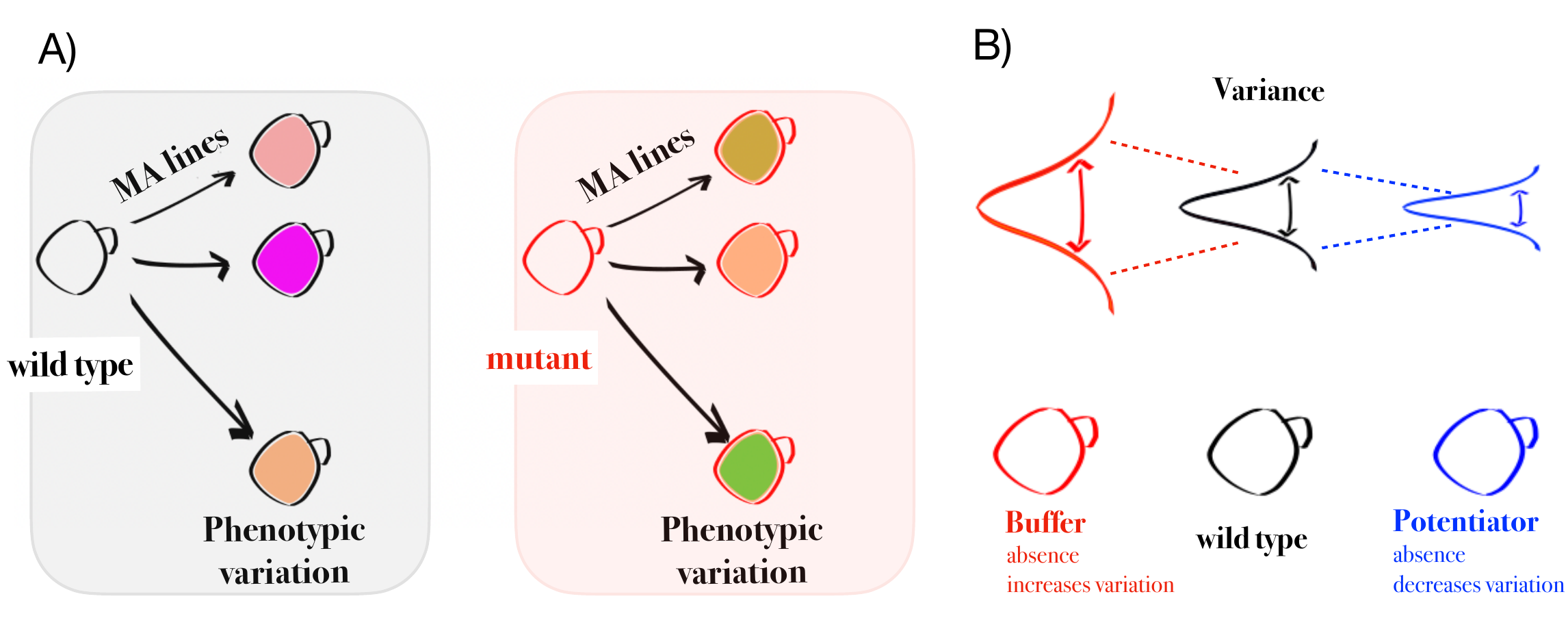}
\caption{Buffers and potentiators of variation. A) Each arrow represents a mutation accumulation (MA) line from a reference budding yeast ({\it Sacharomyces cerevisiae}) genotype (wild type on the left and mutant on the right). Both wild type and mutant experience the same MA trajectories. The filled color indicates the corresponding phenotype of that individual yeast. Given a collection of MA lines, one can calculate the resulting phenotypic variation. B) Mutants that exhibit an increase in variation with respect to the wild type, denoted by the corresponding distribution, are referred to as buffers (shown in red), i.e., the absence of buffers results in an amplified variation. On the other hand, mutants that display a decrease in variation are referred to as potentiators (lack of them reduce variation, as indicated by the blue distribution).}
\label{fig2}
\end{figure}

To illustrate this concept, let's consider the metabolism of the budding yeast {\it Saccharomyces cerevisiae} as a model system~\cite{poyatos_genetic_2020}. A particular MA trajectory will generate mutations altering the activity of the enzymes associated with its metabolism (Figure~\ref{fig2}A). Starting from the wild-type strain, we can generate multiple trajectories and quantify a fixed phenotype of the resulting mutants. Since we have numerous lines, we can compute the variation in this phenotype, such as its variance.

%%%%%%%%%%%%%%%%%%%%%%%%%%%%%%%%%%%%%%%%%%%%%%%%%%%%%%%%%%%%%%%%%%%%%
%%%%%%%%%%%%%%%%%%%%%%%%%%%%%%%%%%%%%%%%%%%%%%%%%%%%%%%%%%%%%%%%%%%%%

Now, let's modify the initial background by introducing a mutation in a single component of the system, such as the enzymes in our example (I represent the change in background by transitioning the contour color from black to red in the yeast cartoon depicted in Figure~\ref{fig2}A). From this new genetic background, we generate the same MA lines and once again measure the phenotypic variation [this could be achieved in practice by introducing the mutation to the selected enzyme in each of the reference MA lines~\cite{richardson_histone_2013}]. In this case, the variance may either increase or decrease compared to the initial background, indicating that certain elements of the system, namely a subset of enzymes, function as buffers or potentiators, respectively, of variation. Please note here that since buffers reduce variation, their corresponding mutants would exhibit higher phenotypic variance. Conversely, in the case of an enzyme working as a potentiator, the opposite effect would be observed (Figure~\ref{fig2}B). 

Moreover, the precise set of variance modulators observed depends on the specific operational regime of the system under investigation. In our example, the functioning of metabolism could be influenced by the specific growth conditions. It is also possible that modifying the operating regime may cause the same agent to change its role, transitioning from a buffer to a potentiator, or vice versa. For instance, TPI1 (Triose-Phosphate Isomerase), an enzyme found in the glycolysis metabolic pathway, serves as a buffer when {\it S. cerevisiae} is growing in a minimal medium or as a potentiator in a rich medium with glycerol as carbon source~\cite{poyatos_genetic_2020}. This highlights the malleability and context-dependent nature of these agents~\cite{geiler-samerotte_selection_2016,richardson_histone_2013,poyatos_genetic_2020}.

%%%%%%%%%%%%%%%%%%%%%%%%%%%%%%%%%%%%%%%%%%%%%%%%%%
\subsection{“Hard” restrictions on variation}
\begin{figure}[bt]
\centering
\includegraphics[width=16cm]{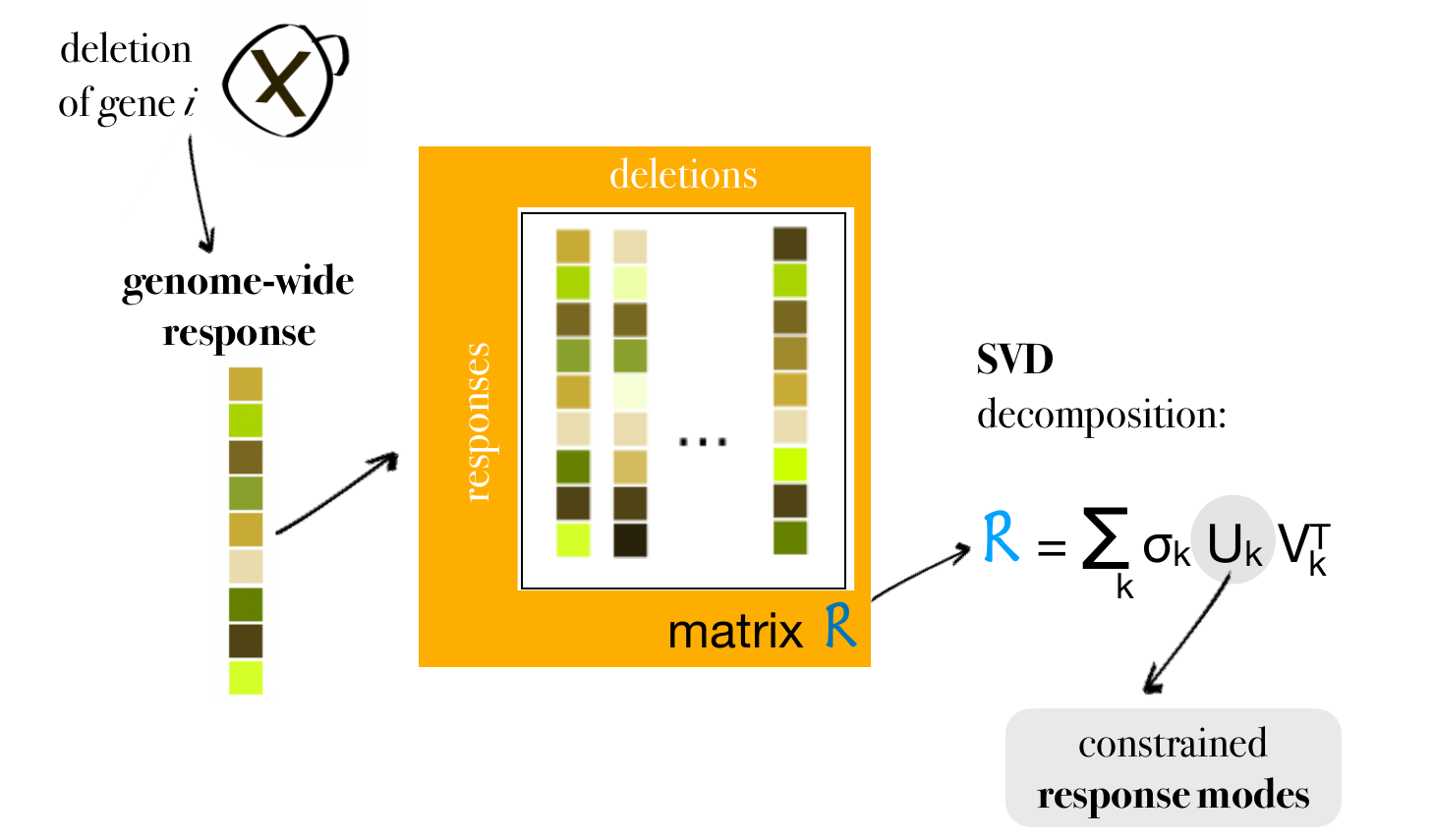}
\caption{Restrictions on variation. The genome responses to single gene deletions were monitored in a collection of budding yeast mutants. This dataset forms a matrix {\large {\tt R}} of transcriptional responses  (rows) versus deletions (columns). To reduce its dimensionality, we employed singular value decomposition (SVD) on {\large {\tt R}} that identifies two sets of orthogonal basis corresponding to the genome responses, $\{U_k\}$, or the deletion assays, $\{V_k\}$. Our hypothesis is that the resulting ${U_k}$ modes obtained through SVD represent constraints on the potential responses of the system, which may lead to suboptimality (main text for further details).}
\label{fig3}
\end{figure}

%%%%%%%%%%%%%%%%%%%%%%%%%%%%%%%%%%%%%%%%%%%%%%%%%%
The second element I present discusses the existence of constraints on variation. The term "constraint" has been a subject of historical discussions in biology, particularly regarding the generation of phenotypic variation, such as developmental constraints ~\cite{smith_developmental_1985}. Nevertheless, my argument proceeds as follows. I consider specifically a dataset obtained with budding yeast, which encompasses hundreds of gene knockouts (representing one-quarter of yeast genes) for which genome-wide mRNA expression was monitored~\cite{kemmeren_large-scale_2014}. This dataset can be represented as a matrix {\tt R} of gene expression {\it vs.} deletions (Figure~\ref{fig3}).

%%%%%%%%%%%%%%%%%%%%%%%%%%%%%%%%%%%%%%%%%%%%%%%%%%%%%%%%%%%%%%%%%%%%%
%%%%%%%%%%%%%%%%%%%%%%%%%%%%%%%%%%%%%%%%%%%%%%%%%%%%%%%%%%%%%%%%%%%%%

To explore the underlying structure of this matrix, we utilize singular value decomposition (SVD), a technique similar to principal component analysis~\cite{strang_linear_2019}. SVD allows us to identify a set of response patterns, $\{ U_k \}$, that we refer to as expression {\it modes}. They represent recurring global expression changes, where groups of genes exhibit coordinated and consistent changes in their expression levels across multiple gene deletion strains~\cite{alter_singular_2000,chagoyen_complex_2019}.

Thus, each transcriptional response can be deconstructed into a limited combination of modes, which I argue reflects the presence of constraints, in this case, on gene expression variation [see~\cite{alba_global_2021} for another recent demonstrative example]. More broadly, the fact that the response is captured by a few modes represents a common property of biological systems, wherein their dynamics can be described by a finite number of degrees of freedom~\cite{eckmann_dimensional_2021}.

An important insight from our research is that these modes might not always yield the most suitable responses. We could envision that if a new condition is encountered, any plastic response would be orchestrated by a combination of a fixed set of expression modes. This could lead to the "erroneous" activation of certain genes, contributing to the observed fitness defects in deletions. This hypothesis was confirmed in our study~\cite{kovacs_suboptimal_2021}. In other words, the presence of stringent restrictions on variation can generate deficient phenotypes. This underscores the significance of considering the interplay between constraints, variation, and fitness.

%%%%%%%%%%%%%%%%%%%%%%%%%%%%%%%%%%%%%%%%%%%%%%
\subsection{“Modular” channeling of variation}
%%%%%%%%%%%%%%%%%%%%%%%%%%%%%%%%%%%%%%%%%%%%%%

My last section addresses the challenge of predicting the fitness consequences of complex, or pleiotropic, mutations. One significant hurdle lies in disentangling the contribution to fitness of the myriad phenotypic effects caused by such mutations, given their sheer diversity. The third principle emphasizes that only a specific subset of the observed variation at the phenotypic level could eventually affect fitness. I explore this phenomenon, which I refer to as "modular" channeling of variation, through the examination of mutations occurring in molecular agents that exhibit high pleiotropy, such as the enzyme RNA polymerase (RNAp)~\cite{yubero_dissecting_2021}. 

In our study, conducted using {\it Escherichia coli} as the experimental model, we examined the impact of specific mutations in the {\it rpoB} gene, which encodes the $\beta$ subunit of RNAp, on gene expression. These mutations commonly lead to resistance to rifampicin, an antibiotic used in the treatment of various bacterial infections~\cite{goldstein_resistance_2014}. Consequently, we quantified various phenotypic traits closely linked to transcriptional efficiency and the alarmone ppGpp, which is a stress signaling pathway in bacteria. Some of these traits are dependent on the environment.

\begin{figure}[bt]
\centering
\includegraphics{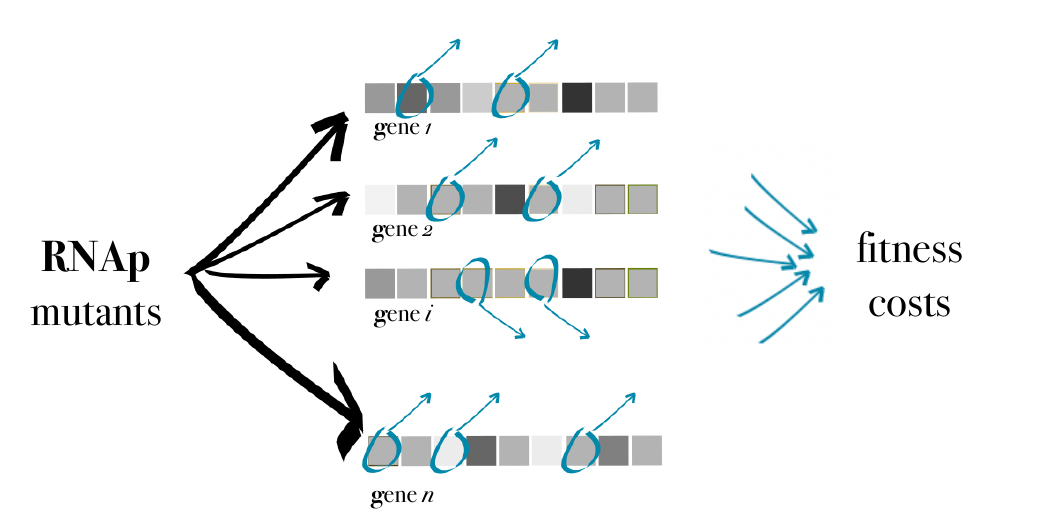}
\caption{Channeling on variation. Mutations in a group of RNAp mutants result in alterations in various characteristics associated with the transcriptional response of constitutive genes (i.e., those not regulated by transcriptional factors). We quantified these features for each gene (represented as "barcodes" in the figure) and observed that only a subset of the phenotypic variation (blue circles/arrows) contributes to fitness variation. This indicates that phenotypic variation is channeled in a modular manner, ultimately shaping the variation observed at the fitness level.}
\label{fig4}
\end{figure}

Each box in Figure~\ref{fig4} represents one such feature and we assessed them by monitoring promoter activity. Importantly, we utilized a set of constitutive genes to measure the expression response to the different RNAp mutations, thereby excluding the interference of additional regulatory elements, such as transcriptional factors. With this data in hand, we developed, for each gene, a model to understand how fitness changes in the mutants as a function of the observed phenotypic features. Interestingly, we found that only a subset of them significantly influences fitness, as measured by the growth rate~\cite{yubero_dissecting_2021}. 

%%%%%%%%%%%%%%%%%%%%%%%%%%%%%%%%%%%%%%%%%%%%%%%%%%%%%%%%%%%%%%%%%%%%
%%%%%%%%%%%%%%%%%%%%%%%%%%%%%%%%%%%%%%%%%%%%%%%%%%%%%%%%%%%%%%%%%%%%

In a broader context, our findings illustrate how specific genetic variations contribute to increased phenotypic diversity. Among the diverse phenotypes represented by the barcodes in Figure~\ref{fig4}, only a subset (depicted by the blue circles) substantially influences fitness. This suggests a model in which pleiotropy may manifest at the phenotypic level, but only a limited number of modified phenotypes ultimately determine fitness. While an earlier study discussed a similar model with "abstract" phenotypes~\cite{kinsler_fitness_2020}, our work enabled us to leverage our understanding of the consequences of mutations in RNAp to quantify a plausible set of phenotypes when changes occur. This observation emphasizes that, while a given mechanistic alteration could lead to many modified phenotypes, only a distinct subset of these alterations appears to be fitness significant, contingent on the environment~\cite{kinsler_fitness_2020,yubero_dissecting_2021}.

%%%%%%%%%%%%%%%%%%%%%%%%%%%%%%%%%%%%%%%%%%%%%%
\section{Conclusions}
%%%%%%%%%%%%%%%%%%%%%%%%%%%%%%%%%%%%%%%%%%%%%%
Three fundamental questions arise from the standard mapping architecture of biology (Figure~\ref{fig1}). First, how did such architecture evolve? 
This question encompasses exploring the different encoding strategies employed in biology, understanding the various developmental mechanisms contributing to phenotypes, and ultimately comprehending how these phenotypes impact fitness. The second question delves into the production of variation at each level of the architecture. This problem also refers to the balance between error generation and tolerance in biology~\cite{frank_evolutionary_2019-1,frank_evolutionary_2019}. Finally, the third problem, which I have examined here, investigates from a functional point of view the mapping of variation from one level to another.

Given the existence of multi-map variation, my focus is to identify generic features that underlie the transformation of variation across different levels. The {\it first principle} highlights the existence of genetic elements with the ability to alter the conversion of genetic into phenotypic variation~\cite{rutherford_hsp90_1998,bergman_evolutionary_2003,schell_modifiers_2016,poyatos_genetic_2020}. These elements have the capacity to amplify (potentiators) or diminish (buffers) the variation at the lower level, both aspects also related to ideas of network controllability [from complex systems~\cite{doyle_robust_1996}] and modifier genes [from genetics~\cite{riordan_peas_2017}]. 

While several molecular agents, such as the heat shock protein HSP90, have been identified as buffers~\cite{rutherford_hsp90_1998,schell_modifiers_2016}, and this buffering mechanism can be attributed to the specific molecular features of the agent [e.g., HSP90 stabilizes unstable signaling proteins for  activation~\cite{rutherford_hsp90_1998}], it is essential to emphasize that this characteristic is not inherently linked to the molecular properties of the element itself. Instead, it might be connected to the role that the element plays within the specific structure under examination~\cite{bergman_evolutionary_2003}, analogous to the enzymes in our metabolic illustration~\cite{poyatos_genetic_2020}. Moreover, this modulatory role is not fixed and can change over time~\cite{geiler-samerotte_selection_2016}. For example, in certain situations, HSP90 has been observed to enhance the effects of genetic variation instead of acting as a buffer~\cite{cowen_hsp90_2005}. Thus, whether an element serves as a buffer or potentiator depends on the specific operational regime of the system under investigation~\cite{poyatos_genetic_2020}.

The existence of constraints, the {\it second principle}, serves to underscore two significant aspects. First, they indicate that biological responses, or behaviours, are both limited and can sometimes result in suboptimal situations~\cite{kovacs_suboptimal_2021}. This observation has been particularly emphasised in discussions surrounding multi-objective optimization in biology, biological trade-offs, and related topics~\cite{shoval_evolutionary_2012}. For example, consider the discussion of phenotypic biases on development. In that context, there was an implicit expectation that morphology could vary in any direction with equal probability~\cite{gerber_not_2014}. Secondly, the functioning of biological systems ultimately operates within a reduced number of dimensions, illustrating the pervasive nature of dimensional reduction, e.g.,~\cite{chagoyen_complex_2019,husain_physical_2020}. 

This becomes increasingly apparent as we have the ability to measure an expanding range of molecular phenotypes, such as through techniques like single-cell RNA sequencing. Despite the vast amount of variation present in these phenotypes, it is intriguing to observe that it can effectively be reduced to a few key dimensions. One could argue that this merely represent the intrinsic sources of variation contributing to the dataset~\cite{strang_linear_2019,eckmann_dimensional_2021}. The question that arises is the biological significance of these sources of variation. For instance, they may correspond to gene expression patterns associated with growth~\cite{chagoyen_complex_2019} or profiles related to responses to environmental changes~\cite{kovacs_suboptimal_2021}. In some other cases, these dimensions can be associated with dynamical attractors~\cite{huang_cell_2005}. 

As {\it third and last principle}, I emphasized the process of channeling variation to the fitness level. It is important to recognise that not all variation at the phenotypic level is necessarily relevant to fitness. This brings us back to the earlier comment I introduce regarding the balance associated with precision. It also evokes the concept of "sloppy" parameters in the context of biological models~\cite{gutenkunst_universally_2007}, where parameters with a wide range of values can yield similar outputs, specifically in relation to fitness.

This analogy highlights the idea that biological systems often exhibit a degree of tolerance for variation that does not substantially affect fitness. Within this context, we encounter the conundrum of understanding the genetic underpinnings of adaptation. Recent studies seem to suggest that only a limited number of phenotypic dimensions contribute to variation~\cite{tenaillon_molecular_2012,venkataram_development_2016}, which may appear contradictory to the high integration observed in organisms, e.g., numerous genome variants impacting multiple traits~\cite{boyle_expanded_2017}. The modular channeling of variation could resolved this conundrum and the trade-off between robustness and adaptability in the face of changing conditions or perturbations~\cite{wagner_robustness_2007,kinsler_fitness_2020}. 

How can we further test these principles? The investigation of variance modulators would necessitate the utilization of MA lines alongside the capability to quantify phenotypes on a large scale, enabling the estimation of the impact of modulators on variance. These prerequisites are met in {\it S. cerevisiae}, where cell morphology serves as the phenotype~\cite{geiler-samerotte_selection_2016}, and can certainly be extended to other experimental models, such as~\cite{hughes_dnaj_2019}. The study of constraints can be conducted in any experiment where numerous molecular or phenotypic features are measurable. In this regard, I anticipate an increasing number of studies emphasizing this concept. For example, a highly constrained set of wing phenotypes was recently characterized using the {\it Drosophila} wing model~\cite{alba_global_2021}. Lastly, the examination of the modular channeling of variation will benefit from the recent methods for mapping high-dimensional genotype-fitness relationships in a more precise and context-dependent manner~\cite{bakerlee_dynamics_2021}.

A final aspect concerns the origins of these principles. Following the spirit of Tinbergen's four questions (mechanism, development, function, and evolutionary history)~\cite{bolhuis_aims_2009}, the aforementioned principles can be associated with the first three questions, but not the fourth. Exploring the selective or neutral processes that give rise to, for example, a molecule functioning as a malleable modulator of variation in a given condition is beyond the scope of this note. However, it presents an intriguing problem that warrants further investigation~\cite{lynch_evolutionary_2012}. Recent discussions on how evolution contributes to this reduction in dimensionality are particularly relevant and merit consideration in this context~\cite{eckmann_dimensional_2021}. Indeed, one could argue that all the above principles emerge within the general context of the low dimensionality of biology.

While I have primarily illustrated these three principles using my own works, it is not meant to suggest that these works hold particular significance over others, as they do not [see, for instance, the very interesting special issues and comments here~\cite{debat_canalization_2019,manrubia_genotypes_2021}]. Rather, they are examples with which I am intimately familiar, and through detailed study of these cases, I have been able to develop and extrapolate the proposed principles. My intent is that this commentary stimulates a more thoughtful discussion of how we can unravel the "organized complexity"~\cite{Weaver_1948} of biology in the 21st century. A deeper understanding of the underlying principles that govern multi-map variation and its impact on biological systems should contribute to that end.

%\begin{epigraph}{Albert Einstein}
%Anyone who has never made a mistake has never tried anything new.
%\end{epigraph}

\section*{acknowledgements}
Most of this manuscript has been developed as a result of the workshop entitled "Theory of plasticity and robustness in natural and artificial systems" held at the Collaboratorium in Barcelona, from July 3 to 5, 2023 and sponsored by the LifeHUB- CSIC, with the aim of promoting open and collaborative debate among researchers from different disciplines. I thank all the participants of this workshop for discussions and Fernando Casares, Ramón Diaz-Uriarte, Juan Rivas-Santisteban and Álvaro Sánchez  for comments on an earlier version of the manuscript.  

%\section*{conflict of interest}
%You may be asked to provide a conflict of interest statement during the submission process. Please check the journal's author guidelines for details on what to include in this section. Please ensure you liaise with all co-authors to confirm agreement with the final statement.

\section{Competing interests}
The authors declare no competing interests

\section{Data Availability} 
No datasets were analyzed.

%===============================%
%========= REFERENCES ==========%
\bibliography{bibliography}
\label{section:references}

\end{document}